\documentclass[submission,copyright,creativecommons]{eptcs}
\providecommand{\event}{EXPRESS/SOS 2025}

\usepackage{iftex}

\ifpdf
  \usepackage{underscore}         
  \usepackage[T1]{fontenc}        
\else
  \usepackage{breakurl}           
\fi

\usepackage{graphicx}
\usepackage{url}
\usepackage{booktabs} 
\usepackage{xspace}
\usepackage{amsmath}
\usepackage{amssymb}
\usepackage{tikz}
\usepackage{fontawesome}
\usepackage[nameinlink,capitalise]{cleveref}
\usepackage[most]{tcolorbox}
\usepackage{amsthm}
\usepackage{float}
\usepackage{listings}
\usepackage[normalem]{ulem}
\usepackage[textwidth=22.0mm,textsize=scriptsize]{todonotes}
\usepackage{dirtree}

\newtheorem{example}{Example}[section]

\pgfdeclarelayer{background}
\pgfdeclarelayer{foreground}
\pgfsetlayers{background,main,foreground}

\newcommand{\bgtext}[1]{%
\bgroup\markoverwith {\textcolor{#1}{\rule[-0.5ex]{2pt}{11pt}}}\ULon}

\usetikzlibrary{arrows.meta}
\usetikzlibrary{fit, positioning, backgrounds, matrix, calc}

\tikzstyle{agent}=[inner sep=2pt, minimum height=5mm, anchor=south,rounded corners=2pt,
                draw=black,font=\scriptsize\sf\color{ptcolor}, align=center]
\tikzstyle{label}=[inner sep=0pt, above=1mm, font=\scriptsize\sf\color{mscolor}]

\definecolor{dkgreen}{rgb}{0,0.6,0}
\definecolor{gray}{rgb}{0.5,0.5,0.5}
\definecolor{mauve}{rgb}{0.58,0,0.82}
\definecolor{codegreen}{rgb}{0,0.6,0}
\definecolor{codegray}{rgb}{0.5,0.5,0.5}
\definecolor{codepurple}{rgb}{0.58,0,0.82}
\definecolor{backcolour}{rgb}{0.95,0.95,0.92}
\definecolor{orange}{RGB}{255,127,0}

\lstdefinestyle{scala}{
    language=Scala,
    morekeywords={
        abstract,case,catch,class,def,do,else,extends,false,final,
        finally,for,forSome,if,implicit,import,lazy,match,new,null,
        object,override,package,private,protected,return,sealed,
        super,this,throw,trait,true,try,type,val,var,while,with,yield,end,then,using
    },
    morekeywords=[2]{Int,Double,Float,Boolean,Char,String,Unit,Any,AnyVal,AnyRef,Nothing,Null,
    Stx,WidgetInfo,Setting,Global,ChannelQueue,Participant,Label,Local,Action,Environment,A},
    morekeywords=[3]{Option,Some,None,Either,Left,Right,List,Seq,Vector,Map,Set,Iterable,Queue},
    sensitive=true,
    morecomment=[l]{//},
    morecomment=[s]{/*}{*/},
    morestring=[b]",
    keywordstyle=\color{blue}\bfseries,
    keywordstyle=[2]\color{teal},
    keywordstyle=[3]\color{red!50!black},
    commentstyle = \color{lime!50!black},
    stringstyle=\color{red!70!black},
    basicstyle=\ttfamily\footnotesize,
    numbers=left,
    numberstyle=\tiny\color{gray},
    numbersep=8pt,
    stepnumber=1,
    tabsize=2,
    breaklines=true,
    showstringspaces=false,
    frame=single,
    captionpos=b,
    aboveskip=2mm,
    belowskip=2mm,
    backgroundcolor=\color{backcolour}
}

\definecolor{compset_colour}{RGB}{100, 85,160}
\definecolor{caos_colour}   {RGB}{180, 20, 30}
\definecolor{webgreen}      {rgb}{  0, .5,  0}
\definecolor{opcolor}       {rgb}{0.4,  0,  0}
\definecolor{richpurple}    {rgb}{0.5,0.2,0.7}
\definecolor{deepamber}     {rgb}{0.85,0.4, 0}

\colorlet{ptcolor}{blue}
\colorlet{mscolor}{webgreen}

\newcommand{\compset}{\textcolor{compset_colour}{CoMPSeT}\xspace}
\newcommand{\caos}{\textcolor{caos_colour}{CAOS}\xspace}
\newcommand{\widget}[1]{\textcolor{richpurple}{#1}}
\newcommand{\baseSemantics}[1]{\textcolor{deepamber}{#1}}

\newcommand{\proja}[1]{\mi{{\downharpoonright_{\pt{#1}}}}}

\newcommand{\intr}[3]{\mi{\pt{#1}\to\pt{#2}:\msg{#3}}}

\newcommand{\pt} [1]{\mi{{\textcolor{blue}    {\mathsf{#1}}}}}
\newcommand{\msg}[1]{\mi{{\textcolor{webgreen}{\mathsf{#1}}}}}

\newcommand{\seq}{~\mathbin{;}~}
\newcommand{\parl}{\mathbin{\|}}

\newcommand{\y}{\textcolor{green!40!black}{\checkmark}}
\newcommand{\n}{\textcolor{red!40!black}{$\times$}}
\newcommand{\hd}[1]{\multicolumn{1}{c}{\textbf{\wrap{#1}}}}

\newcommand{\mi}[1]{\ensuremath{\mathit{#1}}\xspace}
\newcommand{\myparagraph}[1]{\medskip\noindent\textbf{#1} }

\newcommand{\wrap}[1]{\begin{tabular}{@{}c@{}}#1\end{tabular}}

\title{\compset: A Framework for Comparing Multiparty Session Types}
\author{Telmo Ribeiro
    \email{telmo.ribeiro@fc.up.pt}
    \institute{Department of Computer Science}
    \institute{Faculty of Sciences, University of Porto\\Portugal}
\and José Proença
    \email{jose.proenca@fc.up.pt}
    \institute{CISTER \& Department of Computer Science}
    \institute{Faculty of Sciences, University of Porto\\Portugal}
\and Mário Florido
    \email{amflorid@fc.up.pt}
    \institute{LIACC \& Department of Computer Science}
    \institute{Faculty of Sciences, University of Porto\\Portugal}
}
\def\titlerunning{\compset: A Framework for Comparing Multiparty Session Types}
\def\authorrunning{T. Ribeiro, J. Proença \& M. Florido}
\begin{document}
\maketitle

\begin{abstract}
Concurrent systems are often complex and difficult to design.
Choreographic languages, such as Multiparty Session Types (MPST), allow the description of global protocols of interactions by capturing valid patterns of interactions between participants.
Many variations of MPST exist, each one with its rather specific features and idiosyncrasies.
Here we propose a tool -- \compset -- that provides clearer insights over different features in existing MPST.
We select a representative set of MPST examples and provide mechanisms to combine different features and to animate and compare the semantics of concrete examples.
\compset is open-source, compiled into JavaScript, and can be directly executed from any browser, becoming useful both for researchers who want to better understand the landscape of MPST and for teachers who want to explain global choreographies.
\end{abstract}

\section{Introduction} \label{sec:introduction}

Communicating systems can be described by a variety of formalisations, often differing in subtle but significant ways -- such as their treatment of concurrency, message ordering, or assumptions about synchrony -- which makes their analysis non-trivial.
These challenges build upon the inherent difficulties in the architecture of such systems, where numerous execution flows and behaviours must be understood to ensure the absence of communication errors~\cite{CorinneChoreographyAutomata2021}.

This paper focuses on \emph{Multiparty Session Types (MPST)}, a typing discipline that guarantees communication safety and liveness in concurrent systems~\cite{Oven2023}, originally formulated by Honda et al.~\cite{MPAsynchronousST2008}.
\emph{Session Types} denote a formalism capable of verifying correctness in concurrent programs, where the well-behaviour of a protocol can be asserted through the well-typedness of its participants.
The \emph{\underline{Multi}party} aspect generalizes the earlier concept of \emph{Binary Session Types} \cite{LanguagePrimitives1998}, which considered communications between only two parties.

\begin{figure}[htb]
    \centering
    \wrap{\begin{tikzpicture} [
    every node/.style = {
        font=\sffamily,
        draw,
        rectangle,
        rounded corners,
        minimum width=2.5cm,
        minimum height=1cm,
        align=center
    },
    >=Latex,
    thisNode/.style={fill=cyan!10},
]
    
\node (global)      at (0, 0)   [thisNode]  {$\mathcal{G}lobal$};
\node (locals)      at (5, 0)   [thisNode]  {$\mathcal{L}ocals$};
\node (processes)   at (10,0)   [thisNode]  {$\mathcal{P}rocesses$};

\draw[->] (global) -- node[above, draw=none, font=\scriptsize]{projected into} (locals);
\draw[->] (locals) -- node[above, draw=none, font=\scriptsize]{type check}     (processes);

\end{tikzpicture}}
    \caption{The classical Multiparty Session Types' framework}
    \label{fig:mpst-classical-framework}
\end{figure}
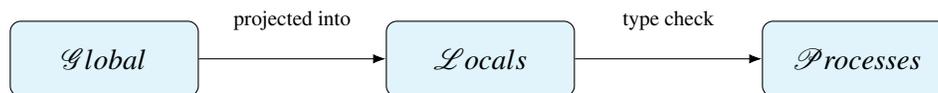

The \emph{classical} MPST framework, illustrated in \cref{fig:mpst-classical-framework}, begins with the specification of a \emph{global type}.
The global type describes the expected communication behaviour of a system from a global perspective, detailing how participants interact and how the session evolves over time, as exemplified in \cref{ex:global-controller-workers}.\footnote{We employ $\to$ to denote the communication of a data type between two participants, while $\seq$ and $\parl$ represent the sequential and parallel composition, respectively.
Additionally, $!$ and $?$ highlight the direction of a communication, marking sending and receiving actions.}

\begin{example} \label{ex:global-controller-workers}
    A possible global type describing a session where a $\pt{controller}$ assigns a task ($\msg{Work}$) to two workers, $\pt{worker_A}$ and $\pt{worker_B}$, in this particular order.
    The workers are then expected to reply with a completion message ($\msg{Done}$) in any order.
    \begin{align*}
        & \pt{controller} \to \pt{worker_A} : \msg{Work} \seq
        \pt{controller} \to \pt{worker_B} : \msg{Work} \seq \\
        &(\pt{worker_A} \to \pt{controller} : \msg{Done} \parl
        \pt{worker_B} \to \pt{controller} : \msg{Done})
    \end{align*}
\end{example}

From the global specification, \emph{local types} -- also known as \emph{session types} -- may be derived via a \emph{projection} operation.
Each local type reflects the perspective of a single participant and contains only the actions in which that participant is involved -- either as sender or as receiver -- similar to the depiction in \cref{ex:locals-controller-workers}.
Local types can then be used to statically type-check processes implementing corresponding participants.

\begin{example} \label{ex:locals-controller-workers}
    The local types derived from the projection of \cref{ex:global-controller-workers}, where $\pt{worker_{AB}}$ is used to reference both $\pt{worker_A}$ and $\pt{worker_B}$, as they share an identical structure.
    \begin{align*}
        L_\pt{controller}  &= \pt{worker_A}!\msg{Work} \seq \pt{worker_B}!\msg{Work} \seq (\pt{worker_A}?\msg{Done} \parl \pt{worker_B}?\msg{Done})\\
        L_\pt{worker_{AB}} &= \pt{controller}?\msg{Work} \seq \pt{controller}!\msg{Done}
    \end{align*}
\end{example}

Formally, let $G$ be a well-formed global type involving participants $\pt{p_1}, \ldots, \pt{p_n}$.
If, for each $1 \leq i \leq n$, there exists a process ${P_i}$ such that $\vdash P_i : (G\proja{p_i})$ -- where $\proja{}$ denotes the projection of $G$ onto participant $\pt{p_i}$ -- then the composed concurrent system $(P_1 \mid ... \mid P_n)$ is guaranteed to be both safe and live~\cite{DynamicMultiroleST2011}.
Here, the projection operation must be a partial function undefined for global types that do not meet the conditions required to ensure these guarantees.

\begin{figure}[htb]
    \centering
    \includegraphics[width=0.8\textwidth]{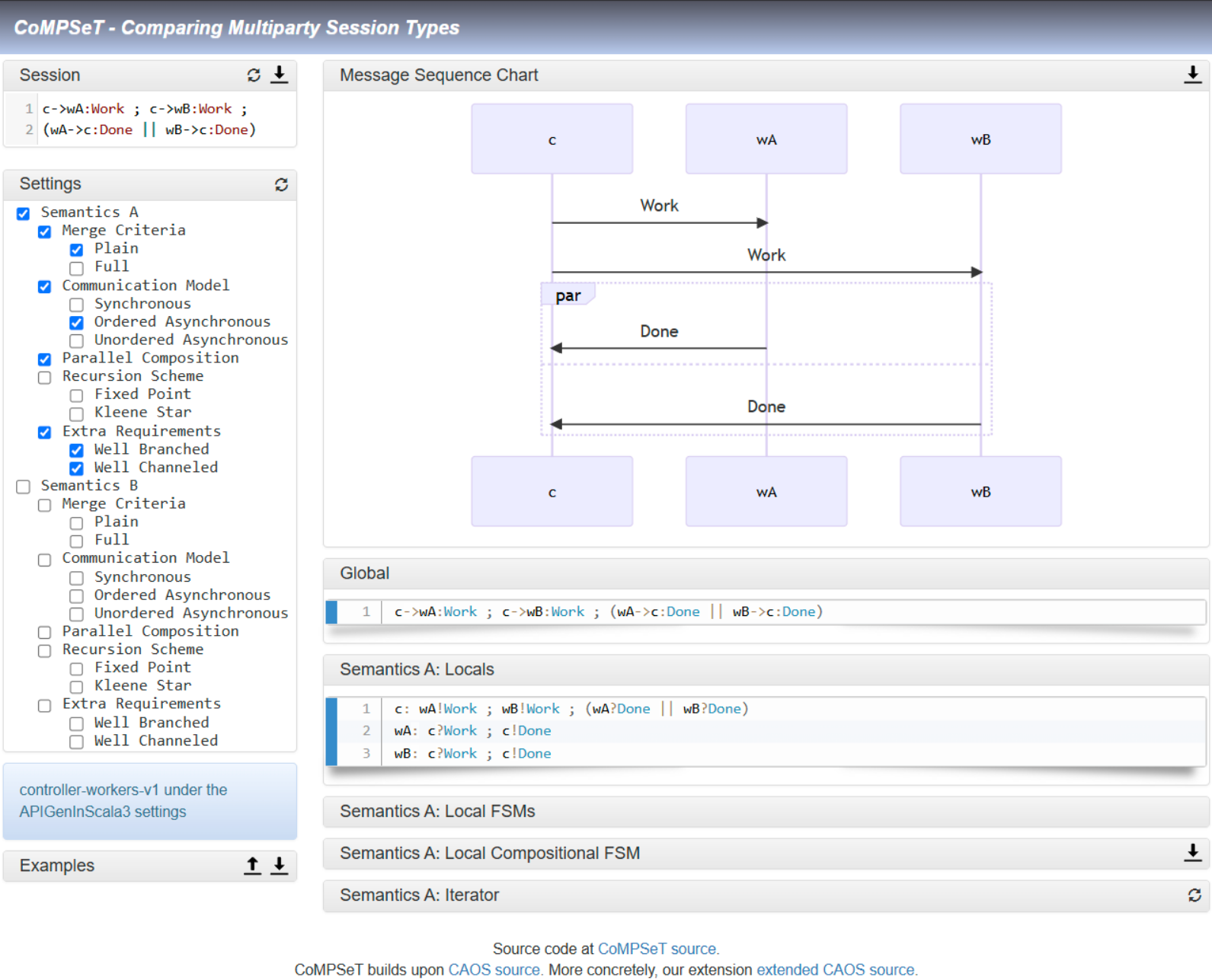}
    \caption{\compset representations for \cref{ex:global-controller-workers} and \cref{ex:locals-controller-workers}, where $\pt{controller}$, $\pt{worker_A}$, and $\pt{worker_B}$ are abbreviated as $\pt{c}$, $\pt{wA}$, and $\pt{wB}$, respectively}
    \label{fig:compset-controller-workers}
\end{figure}

We developed \compset, a tool for comparing MPST sessions and semantics through hands-on experimentation and visualisation.
The core contribution of \compset lies in its ability to support not only the comparison of distinct sessions under the same semantic but also of semantics employing different formalisms, such as synchronous versus asynchronous communication models.
Furthermore, it enables users to configure the underlying semantics according to a selected set of features (see \cref{fig:compset-controller-workers}) and to immediately observe the practical effects of varying formalisation choices.

Our tool pivots \caos~\cite{CAOS2023, CAOS2025}, a framework that enables programmers to test, define, and animate both sessions and structural operational semantics (SOSs) by defining widget (builders) -- visual and/or interactive blocks of extendable functionality -- called upon as functions.
\compset instantiates these builders with appropriate parameters to generate the visual and interactive elements configured according to the user's choice.
Importantly, while \caos is a general-purpose framework for SOSs and does not specifically target MPST, \compset is designed with a modular architecture defining all MPST-specific components, including projections, operational semantics, and well-formedness conditions.
These components are then parametrised against the configurations selected by the user and animated through the \caos framework.

\begin{tcolorbox}[
    colback=yellow!10,
    colframe=yellow!30!black, 
    boxrule=0.5pt,
    arc=2mm,
    left=2mm, right=2mm, top=1mm, bottom=1mm, 
    fontupper=\small,
    title=Remarks on the scope
    ]
We avoid establishing new lemmas and theorems, as our focus is neither on the development of new formulations nor on a comprehensive survey of existing ones, but rather on the introduction of a system capable of effectively comparing MPST variation points.
Although we often provide formalisms to support our notations, they are largely grounded in existing contributions.
\end{tcolorbox}

\myparagraph{Contributions}
Our primary contribution is a prototype open-source tool called \compset, available to be executed online at \url{https://telmoribeiro.github.io/CoMPSeT}.
This tool uses a dedicated input language for specifying global types (see \cref{subsec:GlobalTypes}) and supports configurations over how the language is interpreted, such as choosing between synchronous and asynchronous communication or selecting what constructs are permissible (see \cref{sec:formalisms}).
Users can configure and compare two different semantics within the same session, with additional analysis available through branching bisimulation checkers.
Sessions are visually represented using message sequence charts (MSCs) and projections of the global type, while identifying and reporting errors when projections are undefined.
Finally, it provides semantic animations, either through step-by-step execution or by rendering the entire state space.
As a complementary contribution, we extended the \caos framework to overcome two identified limitations: \textbf{(1)} lack of runtime widget variability and \textbf{(2)} absence of user-driven parametrisation.
These extensions were crucial for supporting the configurability and interactivity of \compset and are made available as an independent fork at \url{https://github.com/TelmoRibeiro/CAOS}.

\myparagraph{Organisation of the paper}
\cref{sec:introduction} introduces the motivation, problem statement, and contributions of this work.
\cref{sec:background} provides an overview of the MPST framework, including definitions of global and local types, projection mechanisms, and semantics for different communication models.
\cref{sec:formalisms} surveys existing formalisms while describing our notions of \emph{feature} and \emph{base semantics}.
\cref{sec:caos} details the extensions made to the \caos framework to support dynamic widget behaviour and user-configurability, enabling the innovative aspects leveraged in \compset.
\cref{sec:compset} defines \compset, describing its interface, features, and use cases through illustrative examples while highlighting how it enables interactive and visual comparisons of different MPST formalisms.
Finally, \cref{sec:conclusion} summarises the main contributions and findings and outlines directions for future research and development.
\section{Multiparty Session Types in a Nutshell} \label{sec:background}

This section serves simultaneously as a gentle introduction to the MPST framework and as the theoretical foundation for understanding the core principles behind our tool.

\subsection{Global Types} \label{subsec:GlobalTypes}

Let $\mathbb{P}$ be the set of all \emph{participants}, ranged over by $\pt{p}$, $\pt{q}$, $\pt{r}$;
let $\mathbb{T}$ be the set of all \emph{data types}, ranged over by $\msg{t}$; and 
let $\mathbb{G}$ be the set of all \emph{global types}, ranged over by $G$.
The syntax of a global type $G$ is given by the following grammar.

\begin{align*}
    G ::= \pt{p} \to \pt{q} : \{\msg{t_i} \seq G_i\}_{1 \leq i \leq n} \mid G_1 \seq G_2 \mid G_1 \parl G_2 \mid \mu X. G \mid X \mid (G)^* \mid skip 
\end{align*}

The syntax and their informal interpretations derive primarily from the descriptions provided by Cledou et al.~\cite{APIGenerationMPSTScala2022}, as well as Jongmans and Proença~\cite{ST4MP2022}, while acknowledging that both developments reference Daniélou and Yoshida~\cite{DynamicMultiroleST2011} as their primary source concerning their own syntax.

\begin{itemize}
    \item $\pt{p} \to \pt{q} : \{\msg{t_i} \seq G_i\}_{1 \leq i \leq n}$ specifies the \emph{communication} of a label $\msg{t_i}$ from the participant \pt{p} to the participant \pt{q}, followed by the global type $G_i$, for some $1 \leq i \leq n$.
    As an additional well-formedness requirement, we stipulate that \textbf{(1)} $\pt{p} \neq \pt{q}$ (i.e., no self-communications) and \textbf{(2)} the labels in $\msg{t_i}$ must be pairwise distinct (i.e., deterministic continuations).
    Moreover, we write $\pt{p} \to \pt{q} : \{\msg{t_i}\}_{1 \leq i \leq n} \seq G$ as a shorthand for $\pt{p} \to \pt{q} : \{\msg{t_i} \seq G\}_{1 \leq i \leq n}$.
   
    \item $G_1 \seq G_2$ specifies the \emph{sequential composition} of $G_1$ and $G_2$.
    
    \item $G_1 \parl G_2$ specifies the \emph{parallel composition} of $G_1$ and $G_2$.
    
    \item $\mu X.G$ and $X$ specify \textbf{(3)} guarded and \textbf{(4)} bounded \emph{recursive protocols} achieved through \emph{fixed point} notation.
    
    \item $(G)^*$ specifies \textbf{(3)} guarded \emph{recursive protocols} achieved through \emph{Kleene star} notation.
    
    \item $skip$ specifies \emph{sequence identity}.
\end{itemize}

Regarding the fixed point notation, we take the equi-recursive viewpoint, not distinguishing between $\mu X.G$ and its unfolding $G[\mu X. G/X]$, as is the usual case within MPST~\cite{MPAsynchronousST2008}.

Furthermore, the constructs adopted are not extensive of the literature, which contains instances such as the \emph{universal quantification} in the work of Daniélou and Yoshida~\cite{DynamicMultiroleST2011}, and \emph{session delegation} established by Bejleri and Yoshida~\cite{SynchronousMPST2008}.

\begin{example} \label{ex:global-recursive-controller-worker}
    The global type below captures a session where the controller delegates a task and the worker responds with a completion note.
    This communication pattern can then be repeated.
    \begin{align*}
        & \mu X. \pt{controller} \to \pt{worker}: \{\\ 
        & \quad \msg{Work} \seq \intr{worker}{controller}{Done} \seq X,\\
        & \quad \msg{Quit}\\
        & \}
    \end{align*}
\end{example}

\subsection{Local Types \& Projections} \label{subsec:LocalTypes&Projections}

Local types are defined by the following grammar.

\begin{align*}
    L ::= \pt{pq}!\{\msg{t_i} \seq L_i\}_{1 \leq i \leq n} \mid \pt{pq}?\{\msg{t_i} \seq L_i\}_{1 \leq i \leq n} \mid L_1 \seq L_2 \mid L_1 \parl L_2 \mid \mu X. L \mid X \mid (L)^* \mid skip
\end{align*}
The informal meaning of the local types is such that:

\begin{itemize}
    \item $\pt{pq}!\{\msg{t_i} \seq L_i\}_{1 \leq i \leq n}$ specifies a \emph{sending} of a label $\msg{t_i}$ from the participant \pt{p} to the participant \pt{q}, followed by the local type $L_i$, for some $1 \leq i \leq n$.
    Moreover, we write $\pt{pq}!\{\msg{t_i}\}_{1 \leq i \leq n} \seq L$ as a shorthand for $\pt{pq}!\{\msg{t_i} \seq L\}_{1 \leq i \leq n}$.

    \item $\pt{pq}?\{\msg{t_i} \seq L_i\}_{1 \leq i \leq n}$ specifies the \emph{reception} of a label $\msg{t_i}$ expected by the participant \pt{q} from the participant \pt{p}, followed by the local type $L_i$, for some $1 \leq i \leq n$.
    Moreover, we write $\pt{pq}?\{\msg{t_i}\}_{1 \leq i \leq n} \seq L$ as a shorthand for $\pt{pq}?\{\msg{t_i} \seq L\}_{1 \leq i \leq n}$.

    \item the remaining constructs -- sequencing, parallel composition, recursion, and skip -- mirror their global type counterparts.
\end{itemize}

The local types are obtained through the \emph{projection} of the global type through each participant.
This notion is formalised in \cref{fig:projection}, which in turn is based upon the definition from Daniélou and Yoshida~\cite{DynamicMultiroleST2011} as well as Yoshida and Gheri~\cite{VeryGentleIntroMPST2020}.

\begin{tcolorbox}[
    colback=yellow!10,
    colframe=yellow!30!black, 
    boxrule=0.5pt,
    arc=2mm,
    left=2mm, right=2mm, top=1mm, bottom=1mm, 
    fontupper=\small,
    title=Remarks on the syntax
    ]
We make the assumption that merging a single local type (see \cref{fig:projection}) returns the same local type, in which case, we omit the braces commonly used to denote several branching continuations.

Whenever it is made clear by the context, we omit the subject of the sending (!) and receiving (?) action.
This approach is not followed in the formulations, which adhere to the literature motifs.
For instance, we omitted $\pt{controller}$ in $L_\pt{controller}$ (see \cref{ex:locals-controller-workers}).

When that is not possible, the notation for communication actions can be made explicit between participants, to avoid ambiguous naming and improve readability.
For example, we use \pt{controller}!\pt{worker}:\msg{Work}, instead of \pt{controllerworker}!\msg{Work}.

For the sake of simplicity, this syntax does not discriminate between payload types and message labels, a characteristic that is not commonly observed in the literature~\cite{MPAsynchronousST2008, MPAsynchronousST2016, SynchronousMPST2008, DynamicMultiroleST2011, ParameterisedMPST2012, LessIsMore2019, GentleIntroMPAsynchronousST2015, VeryGentleIntroMPST2020}.
Rather, following the main sources for the syntax, it uses data types, which may be used to refer interchangeably to either notion.

All the above decisions are transposed to \compset.
\end{tcolorbox}

\begin{figure}[tb]
    \begin{align*}
        &skip \proja{r} = skip\\
        &X \proja{r} = X\\
        &(\mu X. G) \proja{r} = \mu X. (G \proja{r})                                                                                                    &\text{if } \pt{r} \in participants\{G\}\\
        &(\mu X. G) \proja{r} = skip                                                                                                                    &\text{if } \pt{r} \not \in participants\{G\}\\
        &(G)^* \proja{r} = (G \proja{r})^*                                                                                                              &\text{if } \pt{r} \in participants\{G\}\\
        &(G)^* \proja{r} = skip                                                                                                                         &\text{if } \pt{r} \not \in participants\{G\}\\
        &\pt{p} \rightarrow \pt{q} : \{\msg{t_i} \seq G_i\}_{1 \leq i \leq n} \proja{r} = \pt{pq}!\{\msg{t_i} \seq (G_i \proja{r})\}_{1 \leq i \leq n}  &\text{if } \pt{p} = \pt{r} \not = \pt{q}\\
        &\pt{p} \rightarrow \pt{q} : \{\msg{t_i} \seq G_i\}_{1 \leq i \leq n} \proja{r} = \pt{pq}?\{\msg{t_i} \seq (G_i \proja{r})\}_{1 \leq i \leq n}  &\text{if } \pt{p} \not = \pt{r} = \pt{q}\\
        &\pt{p} \rightarrow \pt{q} : \{\msg{t_i} \seq G_1\}_{1 \leq i \leq n} \proja{r} = merge(\{G_i \proja{r}\}_{1 \leq i \leq n})                    &\text{if } \pt{p} \not = \pt{r} \not = \pt{q}\\
        &(G_1 \seq G_2) \proja{r} = (G_1 \proja{r}) \seq (G_2 \proja{r})\\
        &(G_1 \parl G_2) \proja{r} = (G_1 \proja{r}) \parl (G_2 \proja{r})\\
        & \text{undefined}                                                                                                                              &\text{otherwise}
    \end{align*}
    \caption{Projection $G \proja{r}$ of a global type $G\protect\footnotemark$ to a participant \pt{r}, using functions \mi{participants} and \mi{merge}}
    \label{fig:projection}
\end{figure}

The function \mi{participants} returns a set of all participants in a given global type $G$. 
On the other hand, \mi{merge} is a partial function that combines a set of local types $L_i$ into a single one, respecting a strategy that will depend on a \emph{merge criterion}, a point of discussion in \cref{sec:formalisms}.

\begin{tcolorbox}[
    colback=yellow!10,
    colframe=yellow!30!black, 
    boxrule=0.5pt,
    arc=2mm,
    left=2mm, right=2mm, top=1mm, bottom=1mm, 
    fontupper=\small,
    title=Remarks on the Kleene star
    ]
Notably, the Kleene star is not disclosed in the papers referenced for our syntax.  
Instead, the construct was originally introduced by Castagna et al.~\cite{OnGlobalTypesMPS2012}, although it was not fully integrated into their theoretical framework, as it was reduced to a fixed-point operator during projection.
More recently, Jongmans and Proença incorporated support for the Kleene star in their tool implementation, but without an accompanying formal definition in their paper.
In this work, we formalise the rule applied in their implementation.
Importantly, as shown by Charalambides et al.~\cite{ParametrisedConcurrentMPST2012}, the Kleene star can be used to define protocols whose projections are unsafe -- meaning they produce local types that are not compliant with the original global specification.
For instance, the authors present the type $(\pt{a} \to \pt{b}: \msg{m} \seq \pt{b} \to \pt{c}: \msg{m'})^* \seq \pt{c} \to \pt{d}: \msg{m''}$ and explain that $\pt{c}$ cannot determine whether it should wait for $\msg{m'}$ from $\pt{b}$, or skip directly to sending $\msg{m''}$ to $\pt{d}$.
In the same work, the authors propose a Kleene star projectability criterion (\emph{KP}) to ensure that such ambiguity does not arise.
Informally, the criterion requires that all participants in global types of the form $(G_1)^* \seq G_2$ are able to distinguish between $G_1$ and $G_2$.
As such, we assume well-defined projections of the Kleene star to be, additionally, \emph{KP}-consistent. 
\end{tcolorbox}
\footnotetext{We assume \emph{KP}-consistent~\cite{ParametrisedConcurrentMPST2012} global types, following the discussion in \emph{Remarks on the Kleene star}.}

\begin{example} \label{ex:locals-recursive-controller-worker}
    The two local types below capture the projections of the global type from \cref{ex:global-recursive-controller-worker}.
    We present this example before any concrete formulation of \mi{merge}, under the guarantee that all implementations discussed throughout this paper yield the same result for this instance.
    \begin{align*}
        L_\pt{controller} =
        & \mu X.\ \pt{worker}! \{\\
        & \quad \msg{Work} \seq \pt{worker}? \msg{Done} \seq X, \quad \msg{Quit}\\
        \}\\
        L_\pt{worker} =~
        & \mu X. \pt{controller}? \{\\
        & \quad \msg{Work} \seq \pt{controller}! \msg{Done} \seq X, \quad \msg{Quit}\\
        \}
    \end{align*}
\end{example}

\begin{figure}[tb]
    \centering
    \includegraphics[width=0.8\textwidth]{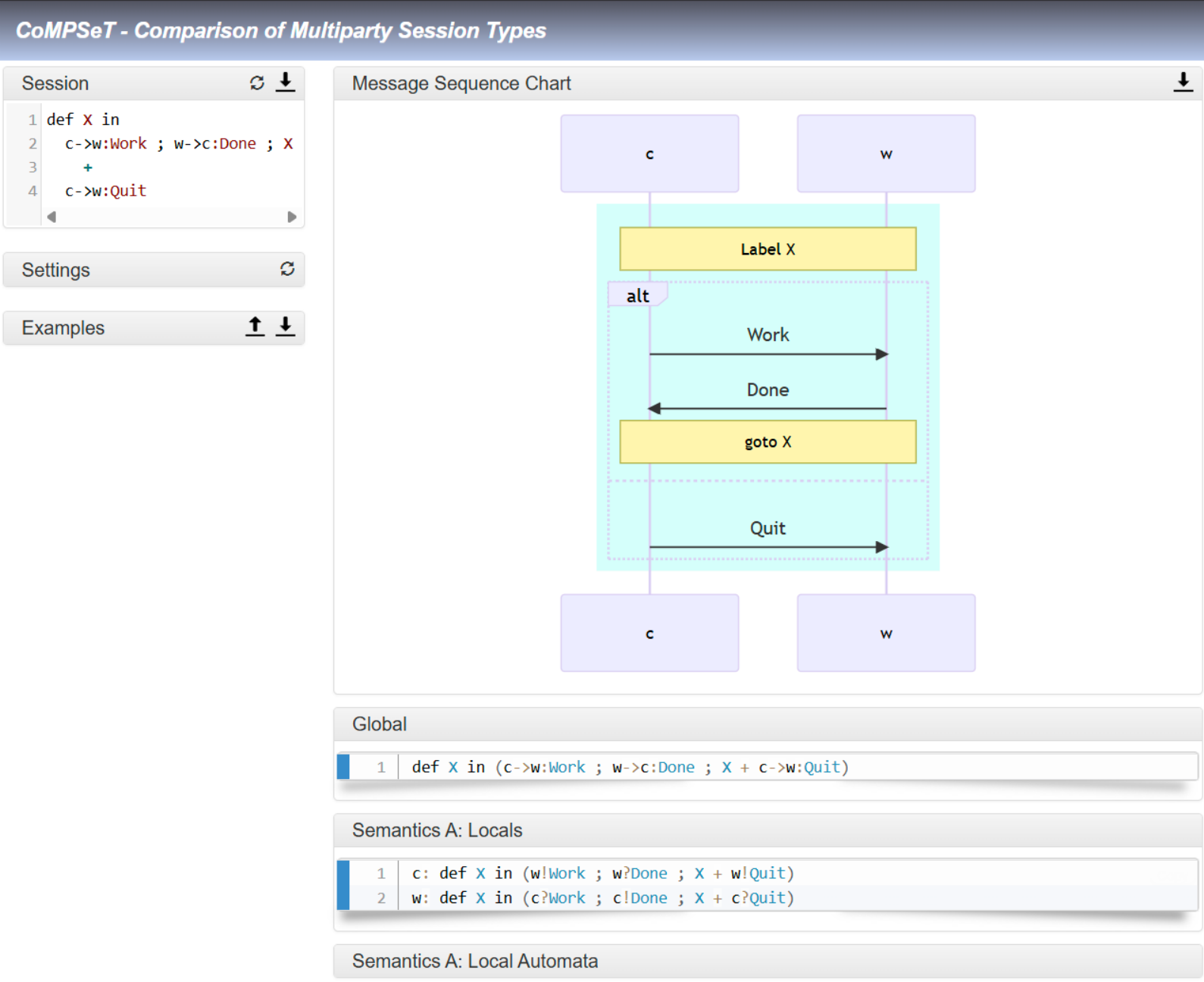}
    \caption{\compset representations for \cref{ex:global-recursive-controller-worker} and \cref{ex:locals-recursive-controller-worker}, where $\pt{controller}$ and $\pt{worker}$ are abbreviated as $\pt{c}$ and $\pt{w}$, respectively}
    \label{fig:compset-recursive-controller-worker}
\end{figure}

Both \cref{ex:global-recursive-controller-worker} and \cref{ex:locals-recursive-controller-worker} are represented in \cref{fig:compset-recursive-controller-worker}, which illustrates their translation into visual widgets in \compset.

\subsection{Running Local Types} \label{subsec:RunningLocalTypes}

We define a \emph{Multiparty Session}, denoted by $M$, as a concurrent composition of local types, written $(L_\pt{p_1} \mid \cdots \mid L_\pt{p_n})$, where each $L_\pt{p}$ is a well-formed local type for some participant $\pt{p} \in M$.
This definition is based on the following considerations: \textbf{(1)} local types can mimic the evolution of well-typed processes, and \textbf{(2)} the employment of semantics over local descriptions is common in choreographic languages beyond MPST.
Our semantics deviate from the standard approach, which typically relies on variants or extensions of $\pi$-calculus to establish reduction rules over processes.

The reduction rules are defined over \emph{configurations} of the form $\langle M,~p\rangle$, consisting of a Multiparty Session $M$ (or, occasionally, a single local type) and a collection $p$ representing possible pending communications.
The realisation of $p$ is established during the definitions of concrete communication rules, when it is assumed to be shared among the remaining reduction rules.
Multisets or first-in first-out (FIFO) queues would be possible realisations of $p$.

As with syntax, \compset supports the full set of reduction rules presented throughout this section, only with minor deviations such as the employment of $env: X \rightarrow L$ -- where $X$ ranges over recursion variables and $L$ over local types -- as mappings for recursion fixed points, which are omitted from the formulations for simplicity. 

We describe the communication rules for three semantics, also omitting structural congruence definitions and shared reduction rules.
Both elements can be found in an extended technical report~\cite{CoMPSeT2025}.

\myparagraph{Synchronous Semantics}
The synchronous communication rule assumes no buffering mechanism, hence $p$ is always empty and absent in the notation.
We assume that $\msg{t_k} \in \bigcup_{i = 1}^{m_i} \msg{t_i} \cap \bigcup_{j = 1}^{m_j} \msg{t_j}$.

\begin{align}
    \langle \pt{pq}! \{ \msg{t_i} \seq L_{1_i} \}_{1 \leq i \leq m_i} \mid
            \pt{pq}? \{ \msg{t_j} \seq L_{2_j} \}_{1 \leq j \leq m_j} \mid
            M 
    \rangle
        &\xrightarrow{\pt{p} \to \pt{q} : \msg{t_k}}
    \langle L_{1_k} \mid L_{2_k} \mid M \rangle \tag{communication}
\end{align}

\myparagraph{Ordered Asynchronous Semantics}
Here, each pair of participants represented simply by $\pt{pq}$ has an unbounded FIFO queue for messages.
A configuration is established by a Multiparty Session $M$ and a buffer $p: (\mathbb{P}\times \mathbb{P}) \to \mathbb{T}^{*}$, mapping each pair $\pt{sender}\text{-}\pt{receiver}$ to a sequence of data types in $\mathbb{T}$.
The main operational rules to evolve a configuration -- send and receive -- are presented below.
Here, $\mathit{ts}$ represents a queue of data types, where $\msg{t_i} \cdot \mathit{ts}$ is used to highlight the prefix element while conversely, $\mathit{ts} \cdot \msg{t_i}$ emphasizes the suffix.

\begin{align}
    \langle \pt{pq}! \{ \msg{t_i} \seq L_i \}_{1 \leq i \leq m} \mid M \,,\, p \cup \{ \pt{pq} \mapsto \mathit{ts} \} \rangle
        &~\xrightarrow{\pt{pq}!\msg{t_k}}~
    \langle L_k \mid M \,,\, p \cup \{ \pt{pq} \mapsto \mathit{ts} {\cdot} \msg{t_k} \} \rangle \tag{send}\\
    \langle \pt{pq}? \{ \msg{t_i} \seq L_i \}_{1 \leq i \leq m} \mid M \,,\, p \cup \{ \pt{pq} \mapsto \msg{t_k} {\cdot} \mathit{ts} \} \rangle
        &~\xrightarrow{\pt{pq}?\msg{t_k}}~
    \langle L_k \,|\, M \,,\, p \cup \{ \pt{pq} \mapsto \mathit{ts} \} \rangle \tag{receive}    
\end{align}

\myparagraph{Unordered Asynchronous Semantics}
Here the configurations are established as pairs of Multiparty Sessions $M$ and a multiset $p \in \mathcal{M}(\mathbb{P} \times \mathbb{P} \times \mathbb{T})$, where $\mathcal{M}(X)$ denotes the set of all finite multisets over the set $X$.
The main operational rules to evolve a configuration are presented below.
We denote by $p \cup (\pt{p}, \pt{q}, \msg{t_i})$ a new multiset achieved by joining the tuple comprising \pt{p},\pt{q}, and \msg{t_i} to an existing multiset~$p$.

\begin{align}
    \langle \pt{pq}! \{ \msg{t_i} \seq L_i \}_{1 \leq i \leq m} \mid M \,,\, p \rangle
        &~\xrightarrow{\pt{pq}!\msg{t_k}}~ 
    \langle L_k \mid M \,,\, p \cup (\pt{p}, \pt{q}, \msg{t_k}) \rangle \tag{send}\\
    \langle \pt{pq}? \{ \msg{t_i} \seq L_i \}_{1 \leq i \leq m} \mid M \,,\, p \cup (\pt{p}, \pt{q}, \msg{t_k}) \rangle
        &~\xrightarrow{\pt{pq}?\msg{t_k}}~ 
    \langle L_k \mid M \,,\, p \rangle \tag{receive}
\end{align}

While \emph{unordered asynchronous} models are atypical in the MPST literature, we include them to showcase the flexibility of the proposed tool, a point elaborated further in the following \cref{sec:formalisms}.
\section{Variations in Multiparty Session Types Formalisms} \label{sec:formalisms}

This section identifies a set of variation points between similar MPST formalisations, regarding their semantics and expressiveness, which we categorise as \emph{features}.
More formally, it defines a feature as a modular aspect in the literature that can vary independently across different semantic implementations.
For example, the realisation of the $merge$ partial function (see \cref{fig:projection}), used to reconcile the projection of branches.
This analysis is inspired by the methodology behind the \emph{essential features} of Bejleri et al.~\cite{ComprehensiveMPST2019}, which compartmentalises structural motifs in MPST.
In contrast, our focus lies specifically on the expressiveness of global and local types while disregarding process-oriented variations.
Additionally, we define \emph{base semantics} as the operational behaviour of our MPST syntax after selecting an explicit set of~features.

The following variation points are considered.

\myparagraph{Merge criteria}
This feature specifies the $merge$ implementation, which handles the projected behaviour of a non-participating role observing branching communication between two others.
We borrow the nomenclature from Scalas and Yoshida~\cite{LessIsMore2019}, thus contemplating the \emph{plain merge} and the \emph{full merge}.
Informally, plain merge is only defined when all branching communications have the \emph{same} continuation, in which case, that continuation is yielded.
The intuition is that if the continuations are the same, the projecting participant does not need to distinguish the branches.
Conversely, full merge -- as introduced by Yoshida et al.\cite{ParameterisedMPST2010,ParameterisedMPST2012} -- extends this notion by allowing distinct yet \emph{compatible}~\cite{LessIsMore2019, VeryGentleIntroMPST2020} communications.

\myparagraph{Communication models}
This feature captures the underlying communication system between participants.
We consider three models:
\textbf{(1)} \emph{synchronous} where senders and receivers communicate in a lock-step; \textbf{(2)} \emph{ordered asynchronous} with an unbounded FIFO queue for pending communications; and \textbf{(3)} \emph{unordered asynchronous} which offers no guarantees on the ordering of pending messages.
The concrete implementations referenced in the literature may differ from those introduced in \cref{subsec:RunningLocalTypes}, which focused on exemplifying how local types could be run natively and introducing a baseline for the comparisons observed in \compset.
For instance, Coppo et al.~\cite{GentleIntroMPAsynchronousST2015} employ a single message queue with additional structural congruence rules to enable reordering, rather than assigning separate queues to each pair \pt{sender\text{-}receiver}.

\myparagraph{Parallel Composition}
This feature captures whether local parallel composition is explicitly supported in the type system, exemplified by Daniélou and Yoshida~\cite{DynamicMultiroleST2011}, Cledou et al.~\cite{APIGenerationMPSTScala2022}, and Jongmans and Proença~\cite{ST4MP2022}.
Notably, some formalisations allow for parallelism at the global type level but not on local types, as evidenced by Bejleri and Yoshida~\cite{SynchronousMPST2008}.
All constructs supported in \compset are done so throughout the complete type system, however, operators concerning only the global types could be established under the new extension discussed in \cref{sec:caos}.

\myparagraph{Recursion Scheme}
This feature describes whether and how recursion is supported -- either as a fixed point construct or via Kleene star notation, representing zero or more repetitions of a term in sequence.
As remarked in \cref{subsec:LocalTypes&Projections}, some systems project the Kleene star as fixed points, forfeiting its native support at the local type level.
A detailed discussion on the expressiveness of different recursion constructs falls outside the scope of this paper.
Yet, recursion in MPST, often blurs the line between expressiveness and syntactic sugar, where we contrast the previous case with the tail-recursive fixed point implementations that could be reduced to Kleene star.
This distinction motivates our decision to implement a local Kleene star with dedicated reduction rules.

\myparagraph{Well-formedness requirements}
This feature regards additional requirements that are conditionally imposed.
For instance, as an additional requirement for the parallel composition, Daniélou and Yoshida~\cite{DynamicMultiroleST2011} require that sub-protocols be \emph{well channelled}, i.e., that their communications do not overlap: $comm(G_1) \cap comm(G_2) = \emptyset$.
Here, $comm : \mathbb{G} \rightarrow 2^{\mathbb{P} \times \mathbb{P} \times \mathbb{T}}$ maps a global type to the set of its communications, represented as triples (\pt{p}, \pt{q}, \msg{t}).

\begin{table}[tb]
\centering
\caption{Features mapping -- \y (present), \n (absent) or N/S (not specified)\protect\footnotemark}
\label{tab:features}
\resizebox{425pt}{!}{%
\begin{tabular}{ lccccc }
    \toprule
    \hd{Paper} & \hd{Merge\\criteria} & \hd{Communication\\model} & \hd{Parallel\\composition} & \hd{Recursion\\scheme} & \hd{Well-formedness\\requirements} \\
    \midrule
    \textbf{\cite{VeryGentleIntroMPST2020}}                       & \wrap{plain \\\& full} & synchronous            & \n  & fixed point                                       &                 \\[2mm]
    \textbf{\cite{GentleIntroMPAsynchronousST2015}}               & plain                  & ordered asynchronous   & \n  & fixed point                                       &                 \\[2mm]
    \textbf{\cite{APIGenerationMPSTScala2022}}                    & plain                  & ordered asynchronous   & \y  & \n                                                & well-channelled \\[2mm]
    \textbf{\cite{ST4MP2022}}                                     & plain                  & ordered asynchronous   & \y  & \wrap{\textbf{Kleene star}\\[-2pt]\& fixed point} & well-channelled \\[2mm]
    \textbf{\cite{RealisabilityPomsetsCommunicatingAutomata2018}} & N/S                    & unordered asynchronous & N/S & N/S                                               & N/S             \\
    \bottomrule
    \end{tabular}%
}
\end{table}

\begin{tcolorbox}[
    colback=yellow!10,
    colframe=yellow!30!black, 
    boxrule=0.5pt,
    arc=2mm,
    left=2mm, right=2mm, top=1mm, bottom=1mm, 
    fontupper=\small,
    title=Remarks on the selected papers
    ]
    The reasoning behind the inclusion of \textbf{\cite{VeryGentleIntroMPST2020}}, \textbf{\cite{GentleIntroMPAsynchronousST2015}}, \textbf{\cite{APIGenerationMPSTScala2022}}, and \textbf{\cite{ST4MP2022}} derives partially from their role as major references while establishing our own formalisms.
    The introductory nature of \textbf{\cite{VeryGentleIntroMPST2020}} and \textbf{\cite{GentleIntroMPAsynchronousST2015}} is shared by the previous sections and allowed for simpler implementations.
    Meanwhile, \textbf{\cite{APIGenerationMPSTScala2022}} and \textbf{\cite{ST4MP2022}} describe tools with design principles shared with \compset, e.g., being accomplished atop \caos.
    In fact, readers familiar with both works can notice how closely our system emulates their original semantics.
    In contrast, \textbf{\cite{RealisabilityPomsetsCommunicatingAutomata2018}} was included because of its unordered asynchronous communications.
    It simultaneously allows for greater variability and stands as motivation for future extensions, since it describes a communication model found in choreographic languages~\cite{RealisabilityPomsetsCommunicatingAutomata2018,RealisabilityPomsets2019} outside the MPST scope.
\end{tcolorbox}
\footnotetext{
Notation used for concrete variation points disregarded in our analysis.
This decision was motivated by the paper describing a choreographic language in which our sole interest was the communication model.
}

\cref{tab:features} illustrates how selected MPST systems combine different features, indicating in the left column a reference to a paper on multiparty communications -- here defined as the conjoined works on MPST and choreographies -- and in the other columns the selection of features used by them.
Notably, in \textbf{\cite{ST4MP2022}} (\emph{ST4MP}) we reference decisions for both the paper and the accompanying implementation (in \textbf{bold}), motivated by the previous discussion on recursion.
\section{Extending \caos} \label{sec:caos}

\caos \cite{CAOS2023, CAOS2025} is defined both as a methodology and a programming framework for computer-aided design of SOSs for formal models.
It supports simultaneous development of semantic foundations and corresponding interactive tools, enabling developers to define reduction rules, use cases structured as examples, and interface elements to visualise and interact with established sessions and operational semantics.
This integration facilitates early detection of incongruences in formalisations, particularly during the modelling and verification of formal semantics.

We selected \caos as the foundation for \compset due to its comprehensive collection of widget builders and its support for visual, interactive means -- a key requirement for the tool.
In particular, we leveraged the following core widgets: \emph{lts} -- to visualise local types or their compositional behaviour under specific communication models; \emph{steps} -- to allow users to interactively compute traces through step-by-step evaluations; and \emph{compareBranchingBisim} -- to determine whether two states under different semantics are branching bisimilar.\footnote{Given the possibility of infinite behaviour, \caos constrains bisimulation checking with a depth bound, set to 100 in \compset.}

\medskip

Despite its flexibility, \caos presented shortcomings for our implementation which other developers may equally face when extending the framework, concerning how to select a set of analysis without overwhelming themselves and/or the user.
For instance, a subtle limitation lies in the way internal configurability is handled.
Although \caos supports multiple forms of semantical equivalence checking -- such as the aforementioned branching bisimulation, but also strong bisimulation and trace equivalence -- its usage is often limited to small sets of semantics.
While it is technically possible to implement parametrised or configurable semantics and compare them through those tools, each configuration needs to be explicitly accounted for, which can quickly lead to scenarios where the number of semantics is just too large to feasibly maintain.

This inflexibility is exemplified by the manner widgets are defined, stopping \caos from dynamically modifying them.

\begin{figure}[H]
    \centering
    \input{src/figures/caos-widgets-signature.tex}
    \caption{Signature for \emph{widgets} in \caos}
    \label{fig:widgets-signature}
\end{figure}

As shown in \cref{fig:widgets-signature}, the selection of widgets is declared as an iterable collection of pairs, each containing the name of a widget and a structure encapsulating its functionality, like the previously discussed \emph{steps}.
This selection of widgets is immutable, reflected by the keyword \emph{val} and by the use of an immutable iterable structure, hence it cannot be updated at runtime to adapt to different configurations.
As a consequence, if one wishes to create a widget that interprets \emph{steps} under synchronous semantics for one example set and asynchronous semantics for another, two distinct widgets must be declared.
Under the current design, this will clutter the web interface with widgets yielding meaningless results or throwing exceptions depending on the input.

To address these limitations, we propose a threefold extension where we: \textbf{(1)} refactored core components of the current implementation to support runtime widget variability; \textbf{(2)} implemented a new configurable input widget -- \widget{Settings} -- whose structure is defined by the developer through a lightweight domain-specific language (DSL) and can be interactively modified by users via checkboxes~\cite{CoMPSeT2025}; and \textbf{(3)} defined an application programming interface (API) over this structure, providing methods for accessing and updating it, alongside general-purpose filters and auxiliary functions~\cite{CoMPSeT2025}.

Those extensions were critical to the innovations observed in \compset.
They enable the programmer to establish widgets that are parametrised by user-selected configurations and which dynamically adapt their behaviour.
More concretely, in \compset, each widget is instantiated by binding its logic to the current state of \widget{Settings}, where we then leverage our modular definitions that capture key aspects of the MPST framework, including global type projection, operational semantics, and well-formedness verification~\cite{CoMPSeT2025}.
\section{Comparing Multiparty Session Types With \compset} \label{sec:compset}

This section presents the \compset tool by describing its applicability, later presenting motivational use cases.
All examples referenced throughout this section -- among others -- are included in the tool~\cite{CoMPSeT2025}.

\subsection{Running \compset} \label{subsec:RunningCompset}

Each setting in the interface (\caos) corresponds directly to a feature (literature) identified in \cref{tab:features}.
Users configure these settings through associated checkboxes, observing the effects on the widgets described.

Two widgets are kept visible and unchanged regardless of configuration: \textbf{(1)} \widget{Message Sequence Chart}, which offers a graphical representation of the session, and \textbf{(2)} \widget{Global} (type), which displays the session specification via text.

\medskip

The configurable settings and their consequent impacts are as follows.

\begin{figure}[tb]
    \centering
    \includegraphics[width=0.8\textwidth]{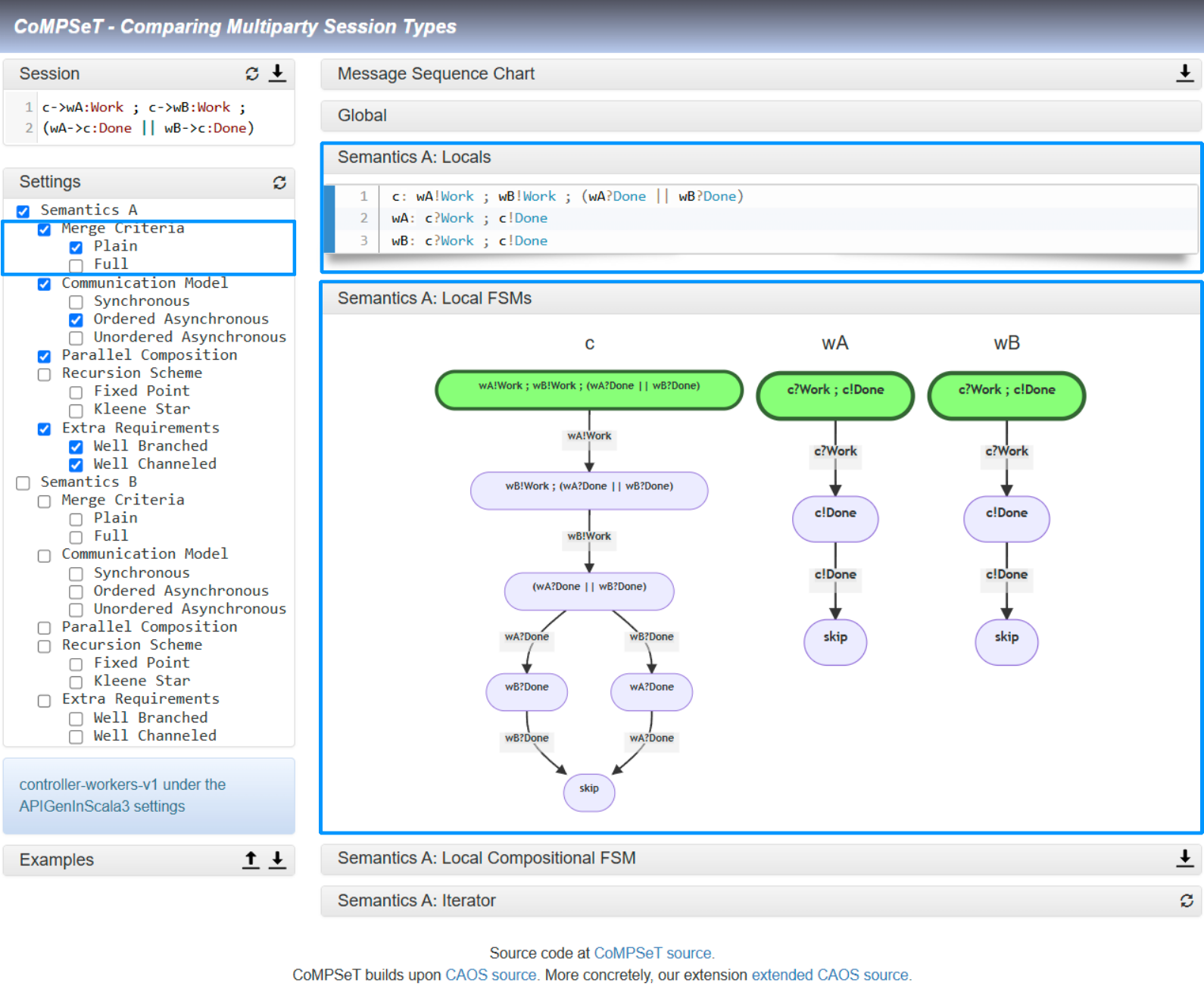}
    \caption{\widget{Locals} and \widget{Local FSMs} for \baseSemantics{APIGenInScala3} (base semantics for \cite{APIGenerationMPSTScala2022}) -- $\pt{controller}\text{-}\pt{workers}$ session from \cref{ex:global-controller-workers}}
    \label{fig:merge-criteria-settings}
\end{figure}

\myparagraph{Merge Criteria}
This setting controls how projection handles branching interactions for non-com\-municating participants.
\compset implements the plain merge following Honda et al.~\cite{MPAsynchronousST2008,MPAsynchronousST2016}, and the full merge according to Dezani-Ciancaglini et al.~\cite{PreciseSubtyping2015}.
These determine whether projection is well-defined, manifested through \widget{Locals} -- a textual description for the local types -- and \widget{Local FSMs} -- its graphical counterpart -- as exemplified by \cref{fig:merge-criteria-settings}.

\begin{figure}[tb]
    \centering
    \includegraphics[width=0.7\textwidth]{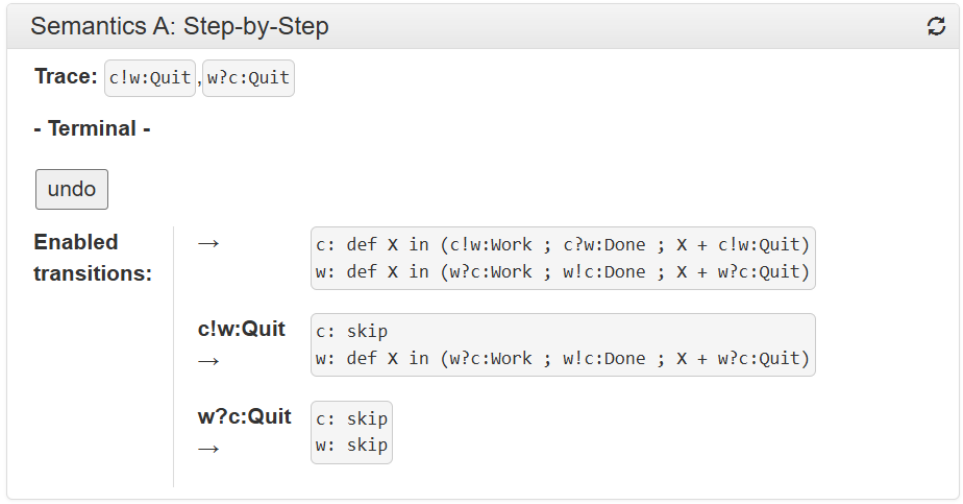}\\
    \includegraphics[width=0.7\textwidth]{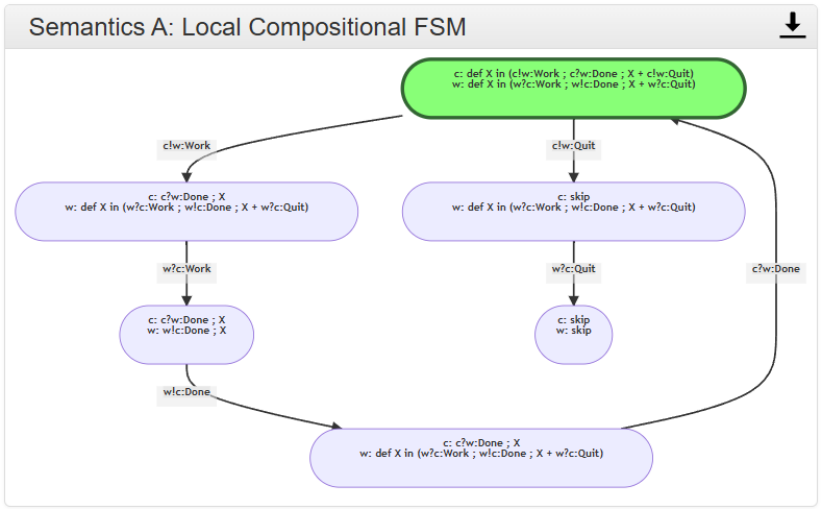}
    \caption{\widget{Step-by-Step} evaluation (top) and \widget{Local Compositional FSM} (bottom) for \baseSemantics{GentleIntroMPAsyncST} (base semantics for \cite{GentleIntroMPAsynchronousST2015}) -- recursive $\pt{controller}\text{-}\pt{worker}$ session from \cref{ex:global-recursive-controller-worker}}
    \label{fig:communication-model-settings}
\end{figure}

\myparagraph{Communication Model}
Users can select between synchronous, ordered asynchronous, or unordered asynchronous communication models, following their descriptions in \cref{subsec:RunningLocalTypes}.
In turn, \widget{Step-by-Step} -- an interactive semantic iterator -- and \widget{Local Compositional FSM} -- a graphical representation of the composed behaviour --  will be rendered on the interface in accordance with \cref{fig:communication-model-settings}.

\begin{figure}[H]
    \centering
    \includegraphics[width=0.35\textwidth]{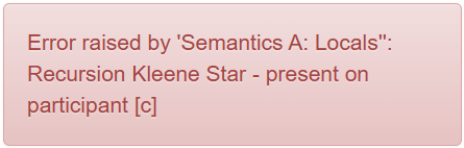}
    \caption{\widget{Check} yielding an error describing the presence of Kleene star recursion for \baseSemantics{VeryGentleIntroMPST} (base semantics for \cite{VeryGentleIntroMPST2020}), using a session described by $(\pt{c} \to \pt{w}: \msg{Work} \seq \pt{w} \to \pt{c}: \msg{Done})^*$}
    \label{fig:check-error-setting}
\end{figure}

\myparagraph{Parallel Composition and Recursion}
These settings control whether constructs like parallel composition and the supported forms of recursion are allowed.
When disabled, sessions containing these operators will raise errors via \widget{Check} -- a widget that runs pre-defined conditions over the system while staying invisible if they do not hold -- as exemplified by \cref{fig:check-error-setting}.
Enabling the corresponding setting suppresses the error.

\myparagraph{Extra Requirements}
This setting enables additional well-formedness checks, such as \emph{well-channelled} (see \cref{sec:formalisms}).
Additionally, \compset also enjoys a relaxed form of branching, expressed as $L_A + L_B$ for local types and $G_A + G_B$ for global types.
To ensure compatibility with classical MPST, we introduce a well-formedness condition named \emph{well-branched}, verifying whether the relaxed form can be rewritten in the canonical syntax.
This decision advances future support for general choreographic languages.
These checks do not suppress errors like the previous settings but add extra syntactic validations.

\subsection{Comparing Semantics} \label{subsec:ComparingSemantics}

\begin{figure}[htbp]
    \centering
    \includegraphics[scale=0.45]{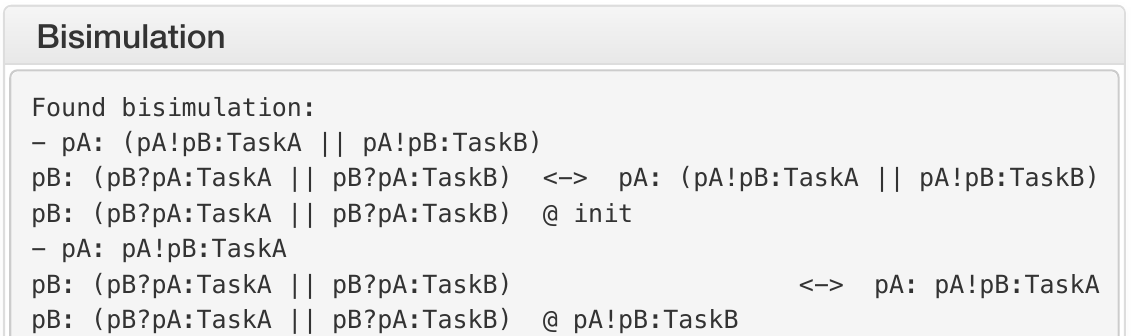}
    \hfill
    \includegraphics[scale=0.45]{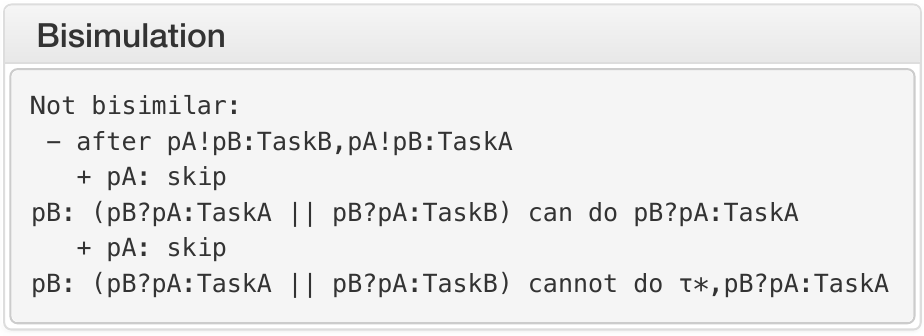}
    \caption{\widget{Bisimulations} comparing the \baseSemantics{APIGenInScala3} semantics either against the \baseSemantics{ST4MP} semantics (left -- only the first seven lines) or against an unordered asynchronous system (right), using a session described by $\pt{pA} \to \pt{pB}: \msg{TaskA} \parl \pt{pA} \to \pt{pB}: \msg{TaskB}$}
    \label{fig:comparative-bisims}
\end{figure}

As a motivational use case, we compare the semantics for \baseSemantics{APIGenInScala3} and \baseSemantics{ST4MP} (base semantics for \cite{ST4MP2022})\footnote{Notably, in \baseSemantics{ST4MP} we adopt the implementation semantics from \cite{ST4MP2022} -- with the Kleene star -- instead of the original formulation, following the discussion of \cref{sec:formalisms} and envisioning greater variability. The formalised version can be experimented upon by selecting instead the \emph{Fixed Point} setting.} which differ only in their recursion treatment, otherwise aligning closely for sessions not foreseeing this construct.
This is evidenced in \cref{fig:comparative-bisims} (left), where \widget{Bisimulation} is a widget describing their behavioural equivalence.

\begin{figure}[htbp]
    \centering
    \includegraphics[width=0.7\textwidth]{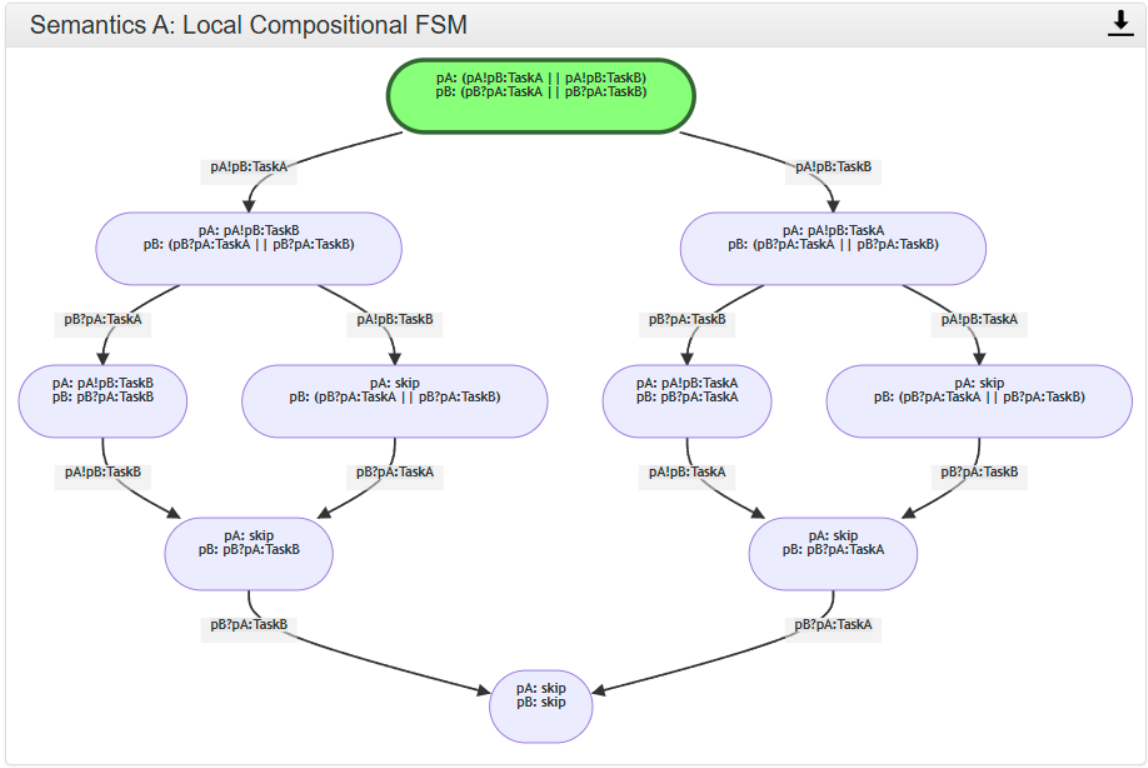}\\
    \includegraphics[width=0.7\textwidth]{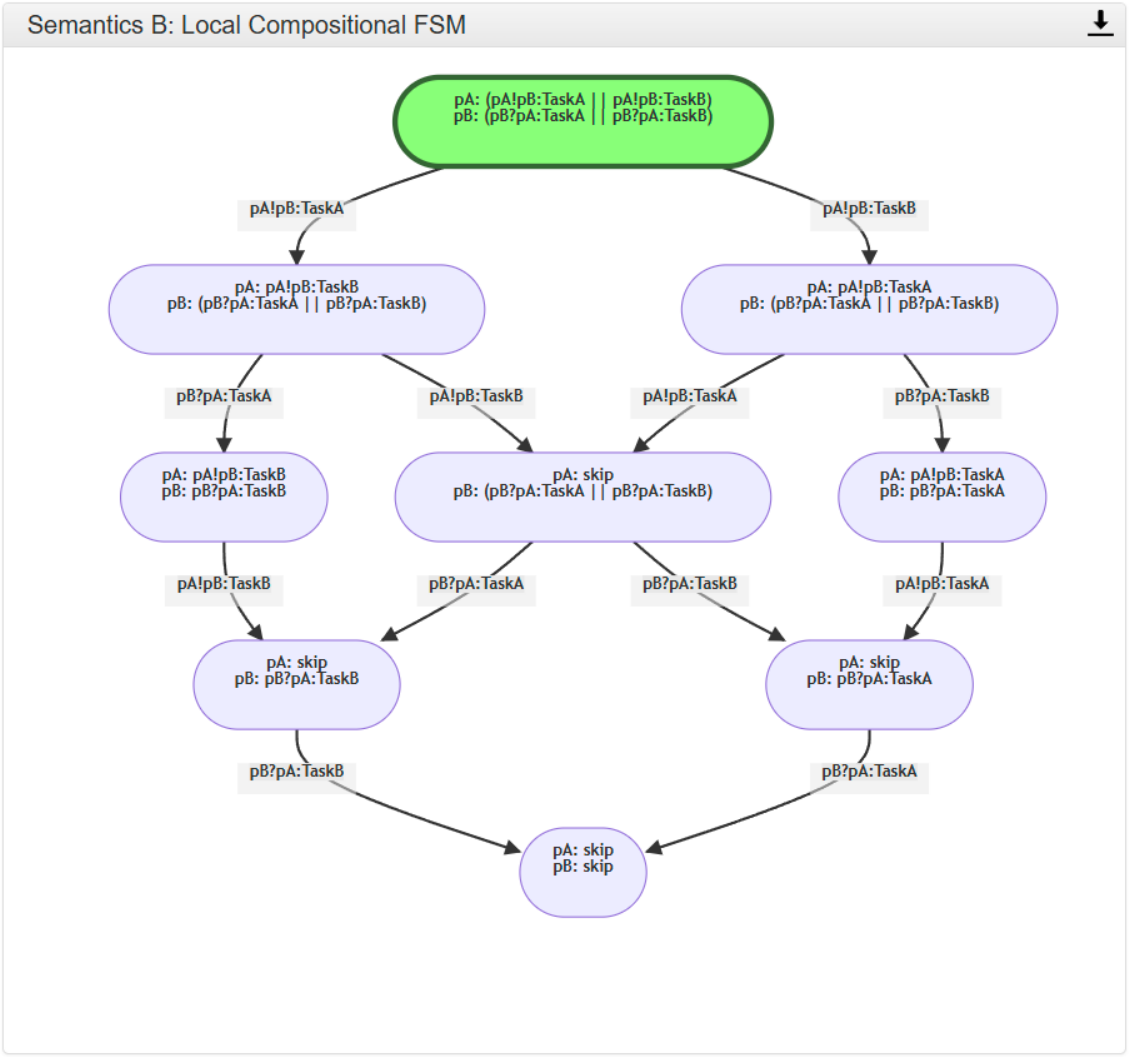}
    \caption{\widget{Local Compositional FSM} for \baseSemantics{APIGenInScala3} (top) and an unordered asynchronous model (bottom) -- simple task delegation session from \cref{fig:comparative-bisims}}
    \label{fig:comparative-local-compositional-fsm}
\end{figure}

However, the previous behavioural equivalence was partially rooted in both semantics sharing the same communication model.
For the same session, a semantic similar with \baseSemantics{ST4MP} yet over an unordered asynchronous model would no longer be bisimilar to \baseSemantics{APIGenInScala3}, as the new communication model does not enforce rules over message delivery.
This nuance is also captured by \compset as illustrated on the right of \cref{fig:comparative-bisims}.
Additionally, the graphical representation for the compositional behaviour (see \cref{fig:comparative-local-compositional-fsm}) further evidences their distinction.

\medskip

\begin{figure}[htbp]
    \centering
    \includegraphics[width=0.8\textwidth]{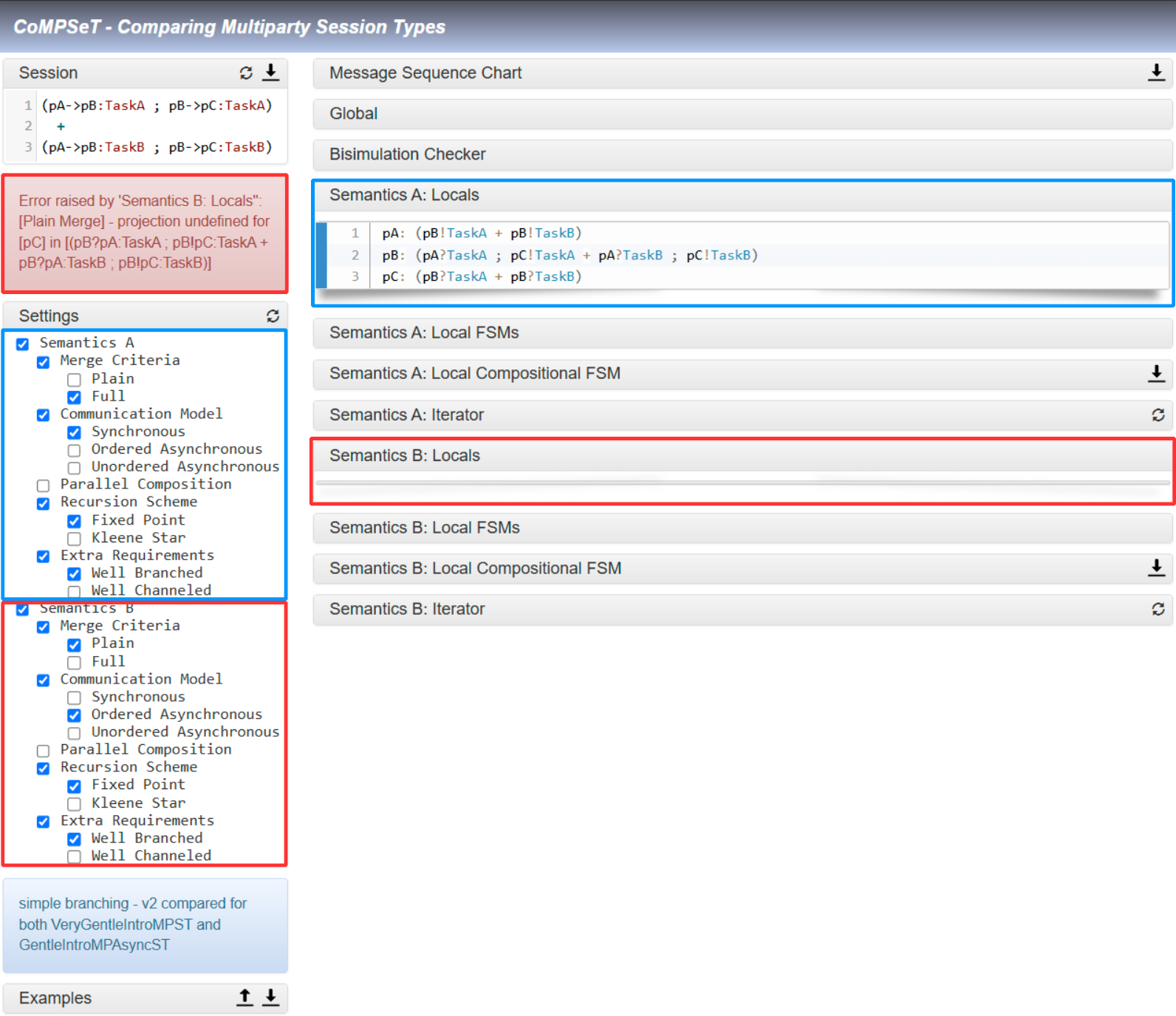}
    \caption{\widget{Locals} for \baseSemantics{VeryGentleIntroMPST} (Semantics A) and \baseSemantics{GentleIntroMPAsyncST} (Semantics B), using the branching session previously described}
    \label{fig:comparative-branching}
\end{figure}

As an additional motivational case, we note how the session defined as $(\pt{pA} \to \pt{pB}: \msg{TaskA} \seq \pt{pB} \to \pt{pC}: \msg{TaskA}) + (\pt{pA} \to \pt{pB}: \msg{TaskB} \seq \pt{pB} \to \pt{pC}: \msg{TaskB})$ is well defined under the full merge, but not under the plain merge, as $\pt{pC}$ would require the same communication action in both branches.
This distinction is captured in \compset -- exemplified in \cref{fig:comparative-branching} -- as \baseSemantics{VeryGentleIntroMPST}, which adopts the full merge, successfully produces the corresponding local type, whereas \baseSemantics{GentleIntroMPAsyncST}, which relies on the plain merge, fails to do so.

\medskip

Readers are encouraged to engage with the examples made accessible online and to configure their own sessions and semantic setups using the presently defined settings.
\section{Conclusion and Future Work} \label{sec:conclusion}

This paper introduced \compset, a novel tool designed to compare Multiparty Session Types (MPST) formalisms through dynamic, user-configurable settings, available at \url{https://telmoribeiro.github.io/CoMPSeT}.

Built atop the \caos framework, \compset leverages its modular widget architecture to define, test, and animate structural operational semantics (SOSs) while exploiting pre-defined notions established by this framework, such as branching bisimulation checkers.
Recognising current limitations in the original \caos framework -- specifically the lack of support for dynamic widget behaviour and runtime configurability -- we extended it to: \textbf{(1)} support runtime variability in widgets; \textbf{(2)} structure configurations through a dedicated domain specific language (DSL); and \textbf{(3)} establish a concise application programming interface (API) that abstracts the definition of compound widget behaviour and facilitates parametrisation.
These extensions are also available as open-source at \url{https://github.com/TelmoRibeiro/CAOS}.

Building on these extensions, \compset benefits from the new \widget{Settings} widget, enabling users to configure the semantics supporting a session and observe, in practice, how different formalism implementations affect system behaviour.
Moreover, it supports side-by-side comparisons of two distinct semantics, providing immediate visual and interactive feedback on their differences.
\compset thus enables researchers and educators to visualise, animate, and compare MPST formalisations, helping to clarify subtle semantic variations within aspects such as communication models, recursion schemes, and merge strategies.

By offering detailed automata visualisation, interactive trace exploration, and bisimulation checking, \compset serves both as an exploratory research platform and as a pedagogical tool for teaching concurrent communication sessions.

\section{Future Work}

Although our tool already supports a range of MPST features, several extensions could broaden its applicability.

\begin{itemize}
    \item \textbf{Additional features --} Further feature assimilation would widen the range of reproducible systems.
    Following the already included form of relaxed branching and unordered asynchronous communications, supplementary adoption of \emph{choreographic features} would extend the scope of the tool allowing for comparisons between different concurrent communication systems;
    \item \textbf{API generation --} MPST tooling often focuses on application programming interface (API) generation \cite{HybridSessionVerification2016, ImplementingMPSTInRust2020, APIGenerationMPSTScala2022, ST4MP2022, Oven2023}, allowing for endpoints implementing different participants to benefit from the guarantees ensured by this typing discipline.
    \compset could be made a foundational layer in a broader pipeline or incorporate this aspect natively, enabling both semantical comparisons and yielding session compliant APIs.
\end{itemize}

By continuing to develop \compset, we aim to lower the barriers to understanding MPST theory and make the impact of different semantic choices more transparent and accessible.

\subsection*{Acknowledgments}
This work is supported by UIDB/00027/2025 of the Artificial Intelligence and Computer Science Laboratory, LIACC, funded by National Funds through FCT/MCTES – \emph{Fundação para a Ciência e a Tecnologia, I.P.} 
(PIDDAC).
This work is also supported by National Funds through FCT/MCTES within the project IBEX, with reference 10.54499/PTDC/CCI-COM/4280/2021; by national funds through FCT/MCTES within the CISTER Research Unit (UIDP/UIDB/04234/2020) and under the project Intelligent Systems Associate Laboratory – LASI (LA/P/0104/2020).


\bibliographystyle{eptcs}
\bibliography{src/references}



\end{document}